\pdfoutput=1

\documentclass{article}

\usepackage{microtype}
\usepackage{graphicx}
\usepackage{booktabs} 

\usepackage{pgfplots}
\usetikzlibrary{intersections}
\usepgfplotslibrary{fillbetween} 
\tikzset{
    axis break gap/.initial=5mm
}
\usepackage{caption}
\usepackage{subcaption}

\usepackage{hyperref}

\usepackage[accepted]{icml2024}

\usepackage{amsmath}
\usepackage{amssymb}
\usepackage{mathtools}
\usepackage{amsthm}

\usepackage[capitalize,noabbrev]{cleveref}

\theoremstyle{plain}

\theoremstyle{definition}

\theoremstyle{remark}

\usepackage{cuted}

\usepackage[textsize=tiny]{todonotes}

\icmltitlerunning{Commute-Time-Optimised Graphs for GNNs}

\begin{document}

\twocolumn[
\icmltitle{Commute-Time-Optimised Graphs for GNNs}

\icmlsetsymbol{equal}{*}

\begin{icmlauthorlist}
\icmlauthor{Igor Sterner}{equal,compsci}
\icmlauthor{Shiye Su}{equal,compsci}
\icmlauthor{Petar Veličković}{compsci,dm}
\end{icmlauthorlist}

\icmlaffiliation{dm}{Google DeepMind}
\icmlaffiliation{compsci}{Department of Computer Science and Technology, University of Cambridge, UK}

\icmlcorrespondingauthor{Petar Veličković}{petarv@google.com}

\icmlkeywords{graph neural networks, rewiring, oversquashing, commute time, priors}

\vskip 0.3in
\editorsListText
\vskip 0.3in
]

\printAffiliationsAndNotice{\icmlEqualContribution} 

\begin{abstract}
We explore graph rewiring methods that optimise commute time.
Recent graph rewiring approaches facilitate long-range interactions in sparse graphs, making such rewirings commute-time-optimal \emph{on average}. 
However, when an expert prior exists on which node pairs should or should not interact, a superior rewiring would favour short commute times between these privileged node pairs.
We construct two synthetic datasets with known priors reflecting realistic settings, and use these to motivate two bespoke rewiring methods that incorporate the known prior.
We investigate the regimes where our rewiring improves test performance on the synthetic datasets.
Finally, we perform a case study on a real-world citation graph to investigate the practical implications of our work.
\end{abstract}

\section{Introduction}

\begin{figure}
    \centering
    \resizebox{\linewidth}{!}{\input{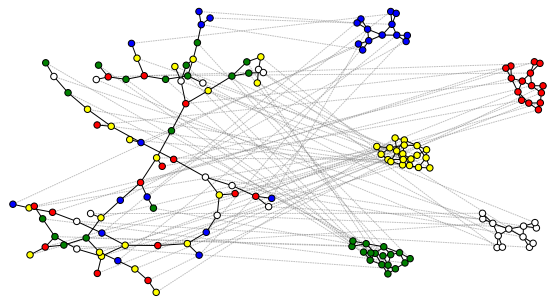}}
    \vspace{-2\baselineskip}
    \caption{A base graph (left) and corresponding Cayley clusters rewiring (right). Each node takes one of four colours, or is uncoloured (white). Our Cayley clusters rewiring sparsely connects nodes of the same colour.}
    \label{fig:data-b-viz}
\end{figure}

Graphs are a natural data structure for a variety of relational networks such as those arising in chemistry, neuroscience, robotics and social sciences \cite{jing2020learning, dezoort2023graph, li2020graph}. 
The dominant paradigm for learning on graph-based data is the graph neural network (GNN) \cite{gori2005new}, which models relational interactions by propagating messages along edges between neighbouring nodes.
Improving GNNs' expressivity has become an area of attention in the machine-learning community \cite{bouritsas2022improving, brody2021attentive}.

The message-passing mechanism suffers fundamentally from the oversquashing phenomenon, whereby bottlenecks in the graph topology result in the aggregation of large numbers of messages onto a small number of fixed-size vector messages.
This makes global information propagation challenging, to the detriment of the GNN's expressivity.
Topological studies on oversquashing show that negative curvature can give rise to these information bottlenecks \cite{topping2022understanding}.
The Hessian of the learning target provides an alternate measure of oversquashing, quantifying the extent to which a GNN can `mix' two nodes and model their pairwise interactions.
This mixing ability is bounded by the GNN's weights and depth \cite{digiovanni2024does}.

In practice, oversquashing would not challenge graph learning if the graph topology was highly aligned to the task, i.e. if nodes that need to interact were always connected.
This motivates \textit{rewiring}, a preprocessing step that constructs an alternate graph for message-passing \cite{alon2020onthe}. 
A suitable rewiring may be found using gradient-based optimisation or reinforcement learning \cite{arnaiz2022diffwire, you2018graph}.
These approaches however suffer from poor scaling due to the large combinatorial search space.
Instead, there has been recent interest in non-parametric approaches, such as Stochastic Discrete Ricci Flow \cite{topping2022understanding}, which improves graph curvature, and expander graph propagation \cite{deac2022expander}, which improves average commute time.
These are more stable, scalable, and theoretically motivated.

Existing non-parametric rewiring methods tend to alleviate oversquashing by optimising for \emph{global} information propagation.
In practice, however, domain experts have \emph{priors} about which nodes need to interact, beyond what is represented in the input graph.
Bottlenecks between two nodes that do not need to interact need not be relieved.
Consider the problem of molecular property prediction, where atoms are modelled as nodes and chemical bonds are modelled as edges. 
Physical laws dictate that all node pairs interact, even distant ones in the original graph.
This is particularly notable for protein molecules, where distant atoms in the primary or secondary structure become proximal in the folded tertiary form. 
Scientists' priors on which of these atoms should interact would offer a valuable headstart for a GNN.

To the best of our knowledge, our work is the first to attempt to incorporate such interaction priors into practical graph rewiring algorithms. 
We propose approaches to integrate input graphs with expert priors on which nodes should mix.
The success of any such approach is contingent on having task-specific priors, and a reasonable means of using said prior in order to aid message passing. 
To disentangle these two effects, we begin by designing synthetic graphs where the priors are perfectly known and focus on developing non-parametric rewiring methods to incorporate them.
As a case study, we then apply our methods to a real-world dataset with a na\"{i}ve prior. 
Our contributions are as follows\footnote{Our code is available at \url{https://github.com/igorsterner/commute-opt-gnn}.}:

\begin{itemize}

\item 
We introduce two new synthetic datasets for graph regression which mirror plausible real-world priors. 

\item
For each of the two datasets, we introduce a non-parametric rewiring to incorporate the prior.

\item
We conduct extensive experiments to evaluate our proposed rewirings.

\item
We conduct a rewiring case study on \texttt{ogbn-arxiv}. 
Our results suggest that rewiring holds the potential for massive performance gains and faster convergence during training. 
\end{itemize}

\section{Expanders for Graph Propagation\label{sec:egp}}

\begin{figure}
    \centering
    \resizebox{0.5\textwidth}{!}{\input{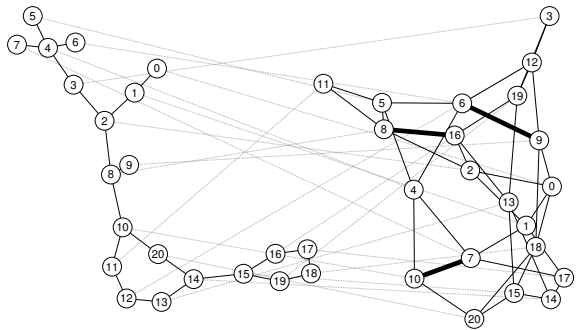}}
    \vspace{-2\baselineskip}
    \caption{A base graph (left) and corresponding trimmed Cayley expander (right). Nodes are randomly allocated to the base graph nodes; the light grey lines show this random allocation. The three thick edges in the Cayley graph are edges which link nodes at distance 5 in the base graph, which is a prior in one of our synthetic datasets.}
    \label{fig:egp-viz}
\end{figure}

A natural way to alleviate the information bottleneck between two nodes is to reduce their shortest path distance or commute time.
Simultaneously, we desire a sparse solution to ensure tractable time and space complexity.
Expander graphs, which are sparse yet low-diameter, satisfy these desiderata.
The latter property ensures high connectivity across all nodes, since the diameter of an expander graph $G$ scales logarithmically with the number of nodes,
\[
\operatorname{diameter} (G) \leq k \log |V(G)|, 
\quad
k \in \mathbb{R^+}
\]
a GNN with $\mathcal{O}(\log(|V(G)|)$ layers suffices to reach all pairwise node interactions.

\citet{deac2022expander} construct sparse expanders using Cayley graphs (see right of Figure~\ref{fig:egp-viz} for a visualisation of a Cayley graph) generated by the special linear group $\operatorname{SL}\left(a, \mathbb{Z}_n\right)$, where $a,n \in \mathbb{Z^+}$.

In our work, we follow their choice of $a=2$ and the generating set
\[
S_n=\left\{\left(\begin{array}{ll}
1 & 1 \\
0 & 1
\end{array}\right),\left(\begin{array}{ll}
1 & 0 \\
1 & 1
\end{array}\right)\right\},
\]
and similarly intervene on standard GNN message passing by propagating every other layer on the Cayley expander, instead of the base graph. 
Expander graphs are less susceptible to oversquashing because the Cheeger constants of a family of expander graphs provably satisfy a constant lower bound $\epsilon > 0$ \cite{alon1986eigenvalues}.
Therefore interleaving in this way retains node--neighbour interactions in the base graph and enables global message propagation in the interleaved graph. 
Hence, \citet{deac2022expander}'s approach can help solve tasks with long-range dependencies.
These ideas have been extended to address higher-order data correlations using hypergraphs \citep{christie2023higherorder}.

Cayley graphs only exist for certain sizes. For our choice of the $\operatorname{SL}\left(2, \mathbb{Z}_n\right)$ group, graph size grows $\mathcal{O}(n^3)$ per:
\begin{equation}\label{equ:num-cayley-nodes}
\left|V\left(\operatorname{Cay}\left(\operatorname{SL}\left(2, \mathbb{Z}_n\right) ; S_n\right)\right)\right|=n^3 \prod_{\text {prime } p \mid n}\left(1-\frac{1}{p^2}\right)
\end{equation}
Following \citet{deac2022expander}, we trim Cayley graphs to arbitrary size $|V|$ by retaining only the first $|V|$ visited nodes in a breadth-first search. 
An example resulting graph is visualised in Figure~\ref{fig:egp-viz}.
However, we note that trimming compromises the graph's expansion properties; keeping the excess nodes as `virtual' zero feature nodes may be preferable \cite{wilson2023cayley}.

A related body of work on graph transformers aims to enable global information propagation.
Graph transformers enable attention-weighted interactions between all node pairs.
But this approach incurs $\mathcal{O}(|V|^2)$ time and space complexity, and can suffer from a large number of redundant messages causing poor out-of-distribution (OOD) generalisation \citet{digiovanni2024does}.
In particular, they show that to allow all nodes to mix in a complete graph, the magnitude of the model's weights need to grow linearly with the size of the graph.
This is due to the fact that complete graphs---such as the ones used in graph Transformers---likely have  redundant messages. In practice, over-specialising on identifying redundant messages on specific training distributions then leads to poor OOD performance on larger graphs.
Exphormers \cite{shirzad2023exphormer} combine these two lines of work by using expanders as attention masks to sparsify the attention layers.

Existing works to date have been concerned with achieving scalable setups that are both sparse and optimise commute time between all nodes.
But sharing information between all nodes is rarely a required characteristic of a graph.
Instead, there will likely exist a subset of the graph nodes which are far apart on the base graph but need to interact to solve the task.
In other words, these approaches are commute time optimal \emph{on average}, but not if there is additional information available on which node pairs should interact most or least.

\section{Synthetic Data Construction}
\label{sec:synth_data_constr}

In this section, we motivate and describe our synthetic datasets.
For simplicity, we choose whole-graph regression tasks, but our ideas generalise to categorical targets and to node- or edge-wise tasks with the same underlying interactions. 
In both datasets, we begin with a graph $G=(V(G), E(G))$, where $V$ and $E$ are the graph nodes and edges.
Node values are sampled from a uniform random variable, i.e. each node takes value sampled from $\mathcal{U}(0,1)$. 

We consider pairwise node interactions via exponential mixing.
Our choice of exponential mixing is motivated by the Hessian-based maximal mixing view of oversquashing \cite{digiovanni2024does}.
Exponential interactions have high mixing and monotonic second derivatives. 
The latter property means it is easy to discern, from features only, which pairwise interactions are more salient for the task.

Specifically, we decompose our graph regression target into three components:

\begin{enumerate}

\item 
A component which depends on nearest-neighbour interactions. 
This justifies the input graph connectivity for the context and task at hand.

\item
A component which depends on pairwise interactions between salient node pairs. 
These interactions are not necessarily captured in the base graph---an expert could inform these through a prior.

\item
A component which depends on pairwise interactions between all remaining nodes. 
This reflects any residual interactions between all pairs of entities in a network not covered by prior components.

\end{enumerate}

By studying different variants of these components, we construct two synthetic datasets.

\paragraph{Data A: Salient pair interactions.}

To concretely motivate this dataset, consider protein property prediction. 
The nearest-neighbour interactions could reflect bonds between consecutive residues on an amino acid chain; the salient node pairs could reflect proximal residues in the folded structure---which may be very distant in the input chain; the all-nodes component could reflect interactions between all residues in the protein due to pairwise electromagnetic force.

We encapsulate this idea in a synthetic dataset where we define salience using a \emph{target} pairwise distance---i.e.,
\begin{align}\label{equ:data-a}
y \ = \
&c_1 \sum_{\textrm{dist}(i,j) = 1} \exp(x_i + x_j) 
+ c_2 \sum_{\textrm{dist}(i,j) = d} \exp(x_i + x_j) \nonumber\\
& + \ c_3 \sum_{\textrm{dist}(i,j) \notin \{1,d\}} \exp(x_i + x_j)
\end{align}
$d \in \mathbb{Z}^+$, and $c_1, c_2, c_3 \in \mathbb{R}_0^+$ are fixed constant parameters, and $x_i$ is a scalar feature given by the node's value. 
The three terms correspond in turn to the three aforementioned components. 
In particular, the second term in Equation~\ref{equ:data-a} represents the interaction of nodes at distance $d$, which is the ``target'' distance at which salient interactions emerge.

\paragraph{Data B: Community interactions.}

This data construction reflects a prior that one or more \emph{subset(s)} of the nodes will need to interact highly in the task.
For example, in the protein property prediction task, there could be a property that depends highly on the structure of one atom type in the input graph.
It is then desirable to minimise commute time between atoms of the same type.

We represent the above atom types as node colours.
Let the set of possible colours be $C$, where $|C| \ll |V|$.
We assign a subset of the nodes $V(c) \subset V$ to each colour $c \in C$, such that each node has at most one colour.
Node features $x_i$ are then a concatenation of node values and a one-hot encoding of $c$ (or zero-vector for uncoloured nodes).

We define the regression target according to:
\begin{equation}\label{equ:data-b}
y = 
c_1 \sum_{\textrm{dist}(i,j) = 1}  \exp(x_i + x_j) + 
c_2 \sum_{c} \sum_{i, j \in V(c)}  \exp(x_i + x_j)
\end{equation}
where $c_1, c_2 \in \mathbb{R}_0^+$ are again fixed constant parameters.
Alongside the same graph-dependent first term as in Equation~\ref{equ:data-a}, 
we introduce a new term here to model mixing between nodes of the same colour.
In Data B, we implicitly set the term describing ``all pairwise interactions'' to zero.
This is simply a design choice, since the colour term already entails a large number of connections.

Finally, we note that though we have described two specific priors, they convey two general, domain-agnostic classes covering pairwise and groupwise interactions respectively.
We believe that successfully addressing these classes allows us to tackle many naturally occurring priors.

\section{Method}

We now propose bespoke rewiring constructions appropriate to the priors built into our two synthetic datasets. As discussed, these rewirings will be introduced by interleaved message passing alongside the base graph.

\paragraph{Data A}

\begin{figure}
    \centering
    \resizebox{0.5\textwidth}{!}{\input{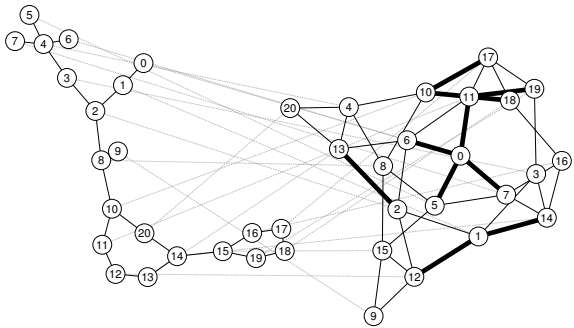}}
    \vspace{-2\baselineskip}
\caption{A base graph (left) and corresponding aligned Cayley expander informed by the prior (right). Compared to random Cayley graph placement (Figure~\ref{fig:egp-viz}), the number of important links captured by the aligned Cayley improves from 3 to 11 (dark edges). The Cayley graphs pictured here and in Figure~\ref{fig:egp-viz} are isomorphic by construction, i.e. we have only changed how the nodes are aligned with the base graph.}
    \label{fig:data-a-viz}
\end{figure}

The base graph naturally supports nearest neighbour interactions (the $c_1$ term).
In addition, we desire a rewiring that can compromise between connecting all distance-$d$ pairs (the $c_2$ term), and allowing all nodes to interact with each other (the $c_3$ term). 
Achieving the latter two objectives independently is straightforward: the former is attained by a graph composed of edges between all distance-$d$ pairs; the latter by Cayley expanders if sparsity is also desired. 

In prior work, nodes in the interleaved graph are randomly matched onto nodes in the base graph.
We propose to align the Cayley expander non-randomly over the nodes of the base graph, such that the salient distance-$d$ edges are more likely to correspond to edges in this aligned Cayley expander. 

This problem distills to the problem of finding the maximum common edge subgraph. As a generalisation of the subgraph isomorphism problem, it is NP-hard. 
We implement a cheap, greedy strategy (Algorithm~\ref{alg:greedy_align}) to identify a correspondence $M$ the two graphs $g_1$ and $g_2$, where $g_1$ and $g_2$ are \texttt{distance-d-pairs} and \texttt{cayley} in our use case. 
An example result of our greedy approximate alignment is visualised in Figure~\ref{fig:data-a-viz}. 
We note, however, that any other alignment scheme (such as backtracking tree-search \cite{McGregor1982BacktrackSA} or reinforcement learning \cite{pmlr-v139-bai21e}) could be used as a drop-in replacement in this step, depending on the computational budget and conviction in the prior. 

\begin{algorithm}[t]
\caption{Greedy alignment of graphs $g_1$ and $g_2$.}
\label{alg:greedy_align}
\begin{algorithmic}

\REQUIRE{Graphs $g_1, g_2$ with the same number of nodes, all nodes initially unassigned}

\STATE $ M = \{\} $

\WHILE{there exist unassigned nodes}

    \STATE $n_1 \gets $ unassigned node in $g_1$
    \STATE $n_2 \gets $ unassigned node in $g_2$
    \STATE $M[n_1] = n_2$

    \WHILE{there exist unassigned nodes in $n_1.\text{neighbours}$}

        \STATE $\tilde{n}_1 \gets $ unassigned node in $n_1$.neighbours
        \STATE $\tilde{n}_2 \gets $ unassigned node in $n_2$.neighbours
        \IF{$\tilde{n}_2$ exists}
		\STATE $M[\tilde{n}_1] = \tilde{n}_2$
        \ENDIF
        
    \ENDWHILE

\ENDWHILE

\STATE \textbf{return} $M$

\end{algorithmic}
\end{algorithm}

\paragraph{Data B}

In addition to the base graph, which retains information about 1-hop neighbours, we aim to find a rewiring that enables nodes of the same colour to interact as easily as possible. 
Meanwhile, we again prefer sparse graphs for computational efficiency and OOD generalisation.

We already have the tool to sparsely connect a graph (or here, subgraph) with optimised commute time: the Cayley expanders.
We therefore generate separate Cayley expanders for nodes of each colour.
We could connect the clusters by introducing sparse random connections between the clusters.
Since here we have strong conviction in our prior (trivially, as it is synthetic data), we do not connect the clusters.\footnote{In our experiments, we found negligible difference from using connected and unconnected Cayley clusters.}
Our Cayley clusters rewiring is visualized in Figure~\ref{fig:data-b-viz}.

\section{Experiment}

\subsection{Experimental Design}

\paragraph{Base graphs}

We construct our synthetic data on base graphs from real-world graph datasets, keeping the graph topology and throwing away node and edge features.
For Data A, base graphs are sourced from ZINC \cite{rafael2016automatic}, which represent real molecules of up to 38 atoms. We sample train graphs of size 20-30 and test graphs of size 30-35.
The OOD design choice allows for a more challenging assessment of the model's ability to generalise, and also probes robustness to distributional shifts that may occur in real-world deployment.
For Data B, we use larger base graphs so that each colour cluster is sufficiently large and non-trivial to sparsely connect.
Graphs are sourced from \texttt{Peptides-struct} of LRGB \cite{dwivedi2022long}.
We sample train graphs of size 75-125 and test graphs of size 125-175.
In each graph, 25-75 nodes are randomly selected and assigned a random colour in $c$.
All remaining nodes are uncoloured.
For both datasets, we sample graphs in bins of 5 nodes to ensure an even spread of graphs of different sizes within the specified boundaries.

\paragraph{Model architecture}

Our architecture is based on the Graph Isomorphism Network (GIN) \cite{xu2018how}. 
This is a message passing GNN with the node update function taking the form of Equation~\ref{equ:gin-update}. $\phi$ is an MLP adapted to include a batch normalization layer before ReLU activation. 
$\epsilon$ is a learnt parameter.
\begin{equation}\label{equ:gin-update}
    \mathbf{h}_u=\phi\left((1+\epsilon) \mathbf{h}_u+\sum_{v \in \mathcal{N}_v} \mathbf{h}_v\right)
\end{equation}

Our final model architecture consists an input linear layer to 8 hidden channels, a stack of five GIN layers, additive pooling and then an output MLP to the regression target.

\paragraph{Learning setup}

A summary of the setup used for experiments on each dataset is shown in Table~\ref{tab:hyperparams}. 

\begin{table}
    \centering
    \caption{Learning setup for the two synthetic datasets} \vspace{0.2cm}
    \begin{tabular}{lll}
    \toprule
         &  \textbf{Data A} & \textbf{Data B} \\ \midrule
         Train graphs & 5000 & 1500 \\
         \quad min size & 20 & 75 \\
         \quad max size & 30 & 125 \\
         \quad batch size & 32 & 32 \\
         Test graphs & 500 & 300 \\
         \quad min size & 30 & 125 \\
         \quad max size & 35 & 175 \\ 
         \quad batch size & 32 & 32 \\ \midrule
         Peak learning rate & 0.0001 & 0.001 \\
         Total epochs & 200 & 200 \\
         Warmup epochs & 50 & 50 \\
         Learning rate decay/epoch & 0.95 & 0.95 \\
         Distance $d$ & 5 & - \\
         Num colours & - & 4 \\
    \bottomrule
    \end{tabular}
    \label{tab:hyperparams}
\end{table}

\paragraph{Rewirers}

For both Data A and Data B, \texttt{base-graph-only} is our baseline; it describes message passing on only the original base graph. 
We also run experiments on \texttt{cayley}, a Cayley expander over all nodes, and \texttt{fully-connected}, a clique over all nodes. 
The bespoke rewirings are: \texttt{aligned-cayley} and \texttt{distance-d-pairs} for Data A; \texttt{cayley-clusters} and a dense variant \texttt{fully-connected-clusters} for Data B.
All rewired graphs interleaved with the base graph in each layer of message propagation.

\paragraph{Metrics}

We measure model performance on the graph regression tasks by mean squared error (MSE) loss.
All evaluation is performed OOD in graph size, as discussed.
We report the means and standard deviations over three runs with randomly-seeded model initialisations.
We compute, for each seed, the final validation set MSE of the rewired approach divided by the that of \texttt{base-graph-only} approach.

\subsection{Results and Discussion}

\paragraph{Data A}

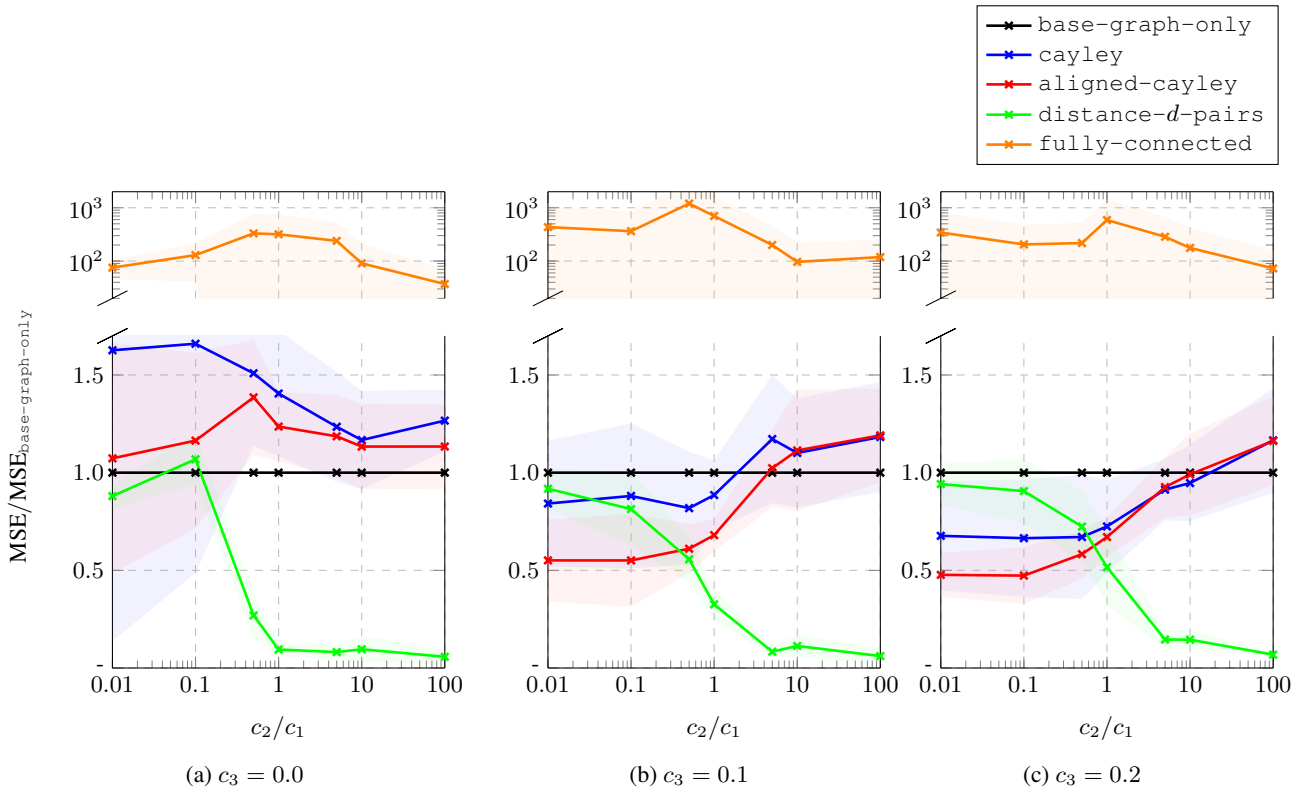
\begin{figure*}[htb]
    \centering
    \vspace{2\baselineskip}
    \begin{small}
    \begin{subfigure}[b]{0.38\textwidth}
        \centering
        \begin{tikzpicture}

\begin{axis}[
    name=bottom axis,   
    width=6cm,
    height=6cm,
    ylabel=$\textrm{MSE}/\textrm{MSE}_{\texttt{base-graph-only}}$,
    xlabel=$c_2/c_1$,
    y label style={at={(0.0,0.7)}},
    xmin = 0.01, xmax = 100,     
    xtick={0.01, 0.1,1,10, 100}, 
    xticklabels={$0.01$, $0.1$, $1$, $10$, $100$}, 
    xmode=log,
    ymin = 0, ymax = 1.7,   
    ytick={0.0, 0.5, 1.0,1.5}, 
    yticklabels={-, $0.5$, $1.0$, $1.5$, $2.0$}, 
    grid = major,
    grid style=dashed,
    axis x line*=bottom,
    legend pos=south east,
]

\addplot[color=black,  mark=x, line width=1pt] coordinates{(0.0100,1.0000)(0.1000,1.0000)(0.5000,1.0000)(1.0000,1.0000)(5.0000,1.0000)(10.0000,1.0000)(100.0000,1.0000)};

\addplot[color=blue,  mark=x, line width=1pt] coordinates{(0.0100,1.6265)(0.1000,1.6600)(0.5000,1.5088)(1.0000,1.4052)(5.0000,1.2348)(10.0000,1.1670)(100.0000,1.2663)};

\addplot[color=red,  mark=x, line width=1pt] coordinates{(0.0100,1.0733)(0.1000,1.1640)(0.5000,1.3855)(1.0000,1.2359)(5.0000,1.1860)(10.0000,1.1336)(100.0000,1.1338)};

\addplot[color=green,  mark=x, line width=1pt] coordinates{(0.0100,0.8800)(0.1000,1.0686)(0.5000,0.2695)(1.0000,0.0947)(5.0000,0.0815)(10.0000,0.0960)(100.0000,0.0575)};


\addplot[name path=only_original_None_top,color=gray!70,draw=none] coordinates {(0.0100,1.0000)(0.1000,1.0000)(0.5000,1.0000)(1.0000,1.0000)(5.0000,1.0000)(10.0000,1.0000)(100.0000,1.0000)};

\addplot[name path=only_original_None_down,color=gray!70,draw=none] coordinates {(0.0100,1.0000)(0.1000,1.0000)(0.5000,1.0000)(1.0000,1.0000)(5.0000,1.0000)(10.0000,1.0000)(100.0000,1.0000)};

\addplot[gray!50,fill opacity=0.1] fill between[of=only_original_None_top and only_original_None_down];

\addplot[name path=interleave_cayley_top,color=blue!70,draw=none] coordinates {(0.0100,3.1154)(0.1000,2.8342)(0.5000,1.8803)(1.0000,1.7273)(5.0000,1.5161)(10.0000,1.4190)(100.0000,1.4244)};

\addplot[name path=interleave_cayley_down,color=blue!70,draw=none] coordinates {(0.0100,0.1377)(0.1000,0.4857)(0.5000,1.1374)(1.0000,1.0831)(5.0000,0.9535)(10.0000,0.9150)(100.0000,1.1083)};

\addplot[blue!50,fill opacity=0.1] fill between[of=interleave_cayley_top and interleave_cayley_down];

\addplot[name path=interleave_aligned_cayley_top,color=red!70,draw=none] coordinates {(0.0100,1.6630)(0.1000,1.6138)(0.5000,1.6807)(1.0000,1.4139)(5.0000,1.3974)(10.0000,1.3528)(100.0000,1.3496)};

\addplot[name path=interleave_aligned_cayley_down,color=red!70,draw=none] coordinates {(0.0100,0.4836)(0.1000,0.7142)(0.5000,1.0903)(1.0000,1.0579)(5.0000,0.9746)(10.0000,0.9145)(100.0000,0.9180)};

\addplot[red!50,fill opacity=0.1] fill between[of=interleave_aligned_cayley_top and interleave_aligned_cayley_down];

\addplot[name path=interleave_distance_d_pairs_top,color=green!70,draw=none] coordinates {(0.0100,0.9446)(0.1000,1.1986)(0.5000,0.3852)(1.0000,0.1288)(5.0000,0.1205)(10.0000,0.1646)(100.0000,0.0910)};

\addplot[name path=interleave_distance_d_pairs_down,color=green!70,draw=none] coordinates {(0.0100,0.8153)(0.1000,0.9386)(0.5000,0.1539)(1.0000,0.0606)(5.0000,0.0425)(10.0000,0.0273)(100.0000,0.0240)};

\addplot[green!50,fill opacity=0.1] fill between[of=interleave_distance_d_pairs_top and interleave_distance_d_pairs_down];




\end{axis}

\begin{axis}[
    at=(bottom axis.north),
    anchor=south, yshift=\pgfkeysvalueof{/tikz/axis break gap},
    width=6cm,
    height=3cm,
    xmin = 0.01, xmax = 100,     
    xtick={0.01, 0.1,1,10, 100}, 
    xticklabels={$0.01$, $0.1$, $1$, $10$, $100$}, 
    xmode=log,
    ymin = 20, ymax = 2000,   
    ytick={10, 100, 1000},          
    ymode=log,
    grid = major,
    grid style=dashed,
    title style={yshift=-1ex, text centered},  
    axis x line*=top,
    legend pos=south east,
    legend cell align=left,
    xticklabel=\empty,
    legend style={at={(2.6,2.0)},anchor=west}, 
    after end axis/.code={
         \draw (rel axis cs:0,0) +(-2mm,-1mm) -- +(2mm,1mm)
              ++(0pt,-\pgfkeysvalueof{/tikz/axis break gap})
              +(-2mm,-1mm) -- +(2mm,1mm)
              (rel axis cs:0,0) +(0mm,0mm) -- +(0mm,0mm)
              ++(0pt,-\pgfkeysvalueof{/tikz/axis break gap})
              +(-2mm,-1mm) -- +(2mm,1mm);
}]

\addplot[color=black,  mark=x, line width=1pt] coordinates{(0.0100,1.0000)(0.1000,1.0000)(0.5000,1.0000)(1.0000,1.0000)(5.0000,1.0000)(10.0000,1.0000)(100.0000,1.0000)};

\addplot[color=blue,  mark=x, line width=1pt] coordinates{(0.0100,1.6265)(0.1000,1.6600)(0.5000,1.5088)(1.0000,1.4052)(5.0000,1.2348)(10.0000,1.1670)(100.0000,1.2663)};

\addplot[color=red,  mark=x, line width=1pt] coordinates{(0.0100,1.0733)(0.1000,1.1640)(0.5000,1.3855)(1.0000,1.2359)(5.0000,1.1860)(10.0000,1.1336)(100.0000,1.1338)};

\addplot[color=green,  mark=x, line width=1pt] coordinates{(0.0100,0.8800)(0.1000,1.0686)(0.5000,0.2695)(1.0000,0.0947)(5.0000,0.0815)(10.0000,0.0960)(100.0000,0.0575)};


\addplot[color=orange,  mark=x, line width=1pt] coordinates{(0.0100,75.4175)(0.1000,129.4203)(0.5000,330.8558)(1.0000,317.9252)(5.0000,239.2933)(10.0000,90.7615)(100.0000,37.1106)};

\addplot[name path=interleave_fully_connected_top,color=orange!70,draw=none] coordinates {(0.0100,104.7773)(0.1000,217.9588)(0.5000,770.2374)(1.0000,757.5368)(5.0000,538.1082)(10.0000,210.5849)(100.0000,68.9161)};

\addplot[name path=interleave_fully_connected_down,color=orange!70,draw=none] coordinates {(0.0100,46.0578)(0.1000,40.8818)(0.5000,0.0001)(1.0000,0.0001)(5.0000,0.0001)(10.0000,0.0001)(100.0000,5.3050)};

\addplot[orange!50,fill opacity=0.1] fill between[of=interleave_fully_connected_top and interleave_fully_connected_down];

\legend{
\texttt{base-graph-only}, 
\texttt{cayley},
\texttt{aligned-cayley},
\texttt{distance-$d$-pairs}, 
\texttt{fully-connected},
}
\end{axis}
\end{tikzpicture}
        \vspace{-2\baselineskip}
        \caption{$c_3=0.0$}
    \end{subfigure}
    \begin{subfigure}[b]{0.3\textwidth}
        \centering
        \begin{tikzpicture}

\begin{axis}[
    name=bottom axis,   
    width=6cm,
    height=6cm,
    xlabel=$c_2/c_1$,
    y label style={at={(0.0,1.0)}},
    xmin = 0.01, xmax = 100,     
    xtick={0.01, 0.1,1,10, 100}, 
    xticklabels={$0.01$, $0.1$, $1$, $10$, $100$}, 
    xmode=log,
    ymin = 0, ymax = 1.7,   
    ytick={0.0, 0.5, 1.0,1.5}, 
    yticklabels={-, $0.5$, $1.0$, $1.5$, $2.0$}, 
    grid = major,
    grid style=dashed,
    axis x line*=bottom,
    legend style={at={(2.4,1.3)},anchor=west}, 
]

\addplot[color=black,  mark=x, line width=1pt] coordinates{(0.0100,1.0000)(0.1000,1.0000)(0.5000,1.0000)(1.0000,1.0000)(5.0000,1.0000)(10.0000,1.0000)(100.0000,1.0000)};

\addplot[color=blue,  mark=x, line width=1pt] coordinates{(0.0100,0.8419)(0.1000,0.8808)(0.5000,0.8185)(1.0000,0.8854)(5.0000,1.1720)(10.0000,1.1008)(100.0000,1.1827)};

\addplot[color=red,  mark=x, line width=1pt] coordinates{(0.0100,0.5509)(0.1000,0.5508)(0.5000,0.6106)(1.0000,0.6800)(5.0000,1.0229)(10.0000,1.1138)(100.0000,1.1910)};

\addplot[color=green,  mark=x, line width=1pt] coordinates{(0.0100,0.9178)(0.1000,0.8137)(0.5000,0.5568)(1.0000,0.3259)(5.0000,0.0836)(10.0000,0.1129)(100.0000,0.0612)};


\addplot[name path=only_original_None_top,color=gray!70,draw=none] coordinates {(0.0100,1.0000)(0.1000,1.0000)(0.5000,1.0000)(1.0000,1.0000)(5.0000,1.0000)(10.0000,1.0000)(100.0000,1.0000)};

\addplot[name path=only_original_None_down,color=gray!70,draw=none] coordinates {(0.0100,1.0000)(0.1000,1.0000)(0.5000,1.0000)(1.0000,1.0000)(5.0000,1.0000)(10.0000,1.0000)(100.0000,1.0000)};

\addplot[gray!50,fill opacity=0.1] fill between[of=only_original_None_top and only_original_None_down];

\addplot[name path=interleave_cayley_top,color=blue!70,draw=none] coordinates {(0.0100,1.1642)(0.1000,1.2521)(0.5000,1.1066)(1.0000,1.0645)(5.0000,1.4997)(10.0000,1.3801)(100.0000,1.4650)};

\addplot[name path=interleave_cayley_down,color=blue!70,draw=none] coordinates {(0.0100,0.5197)(0.1000,0.5096)(0.5000,0.5304)(1.0000,0.7063)(5.0000,0.8444)(10.0000,0.8215)(100.0000,0.9004)};

\addplot[blue!50,fill opacity=0.1] fill between[of=interleave_cayley_top and interleave_cayley_down];

\addplot[name path=interleave_aligned_cayley_top,color=red!70,draw=none] coordinates {(0.0100,0.7616)(0.1000,0.7883)(0.5000,0.7364)(1.0000,0.7635)(5.0000,1.2214)(10.0000,1.4258)(100.0000,1.4313)};

\addplot[name path=interleave_aligned_cayley_down,color=red!70,draw=none] coordinates {(0.0100,0.3402)(0.1000,0.3132)(0.5000,0.4847)(1.0000,0.5966)(5.0000,0.8244)(10.0000,0.8018)(100.0000,0.9506)};

\addplot[red!50,fill opacity=0.1] fill between[of=interleave_aligned_cayley_top and interleave_aligned_cayley_down];

\addplot[name path=interleave_distance_d_pairs_top,color=green!70,draw=none] coordinates {(0.0100,1.0224)(0.1000,1.0033)(0.5000,0.6729)(1.0000,0.4060)(5.0000,0.1060)(10.0000,0.1712)(100.0000,0.1012)};

\addplot[name path=interleave_distance_d_pairs_down,color=green!70,draw=none] coordinates {(0.0100,0.8133)(0.1000,0.6242)(0.5000,0.4408)(1.0000,0.2459)(5.0000,0.0613)(10.0000,0.0546)(100.0000,0.0212)};

\addplot[green!50,fill opacity=0.1] fill between[of=interleave_distance_d_pairs_top and interleave_distance_d_pairs_down];




\end{axis}

\begin{axis}[
    at=(bottom axis.north),
    anchor=south, yshift=\pgfkeysvalueof{/tikz/axis break gap},
    width=6cm,
    height=3cm,
    xmin = 0.01, xmax = 100,     
    xtick={0.01, 0.1,1,10, 100}, 
    xticklabels={$0.01$, $0.1$, $1$, $10$, $100$}, 
    xmode=log,
    ymin = 20, ymax = 2000,   
    ytick={10, 100, 1000},          
    ymode=log,
    grid = major,
    grid style=dashed,
    title style={yshift=-1ex, text centered},  
    axis x line*=top,
    legend pos=south east,
    legend cell align=left,
    xticklabel=\empty,
    legend pos=north east,
    after end axis/.code={
         \draw (rel axis cs:0,0) +(-2mm,-1mm) -- +(2mm,1mm)
              ++(0pt,-\pgfkeysvalueof{/tikz/axis break gap})
              +(-2mm,-1mm) -- +(2mm,1mm)
              (rel axis cs:0,0) +(0mm,0mm) -- +(0mm,0mm)
              ++(0pt,-\pgfkeysvalueof{/tikz/axis break gap})
              +(-2mm,-1mm) -- +(2mm,1mm);
}]


\addplot[color=black,  mark=x, line width=1pt] coordinates{(0.0100,1.0000)(0.1000,1.0000)(0.5000,1.0000)(1.0000,1.0000)(5.0000,1.0000)(10.0000,1.0000)(100.0000,1.0000)};

\addplot[color=blue,  mark=x, line width=1pt] coordinates{(0.0100,0.8419)(0.1000,0.8808)(0.5000,0.8185)(1.0000,0.8854)(5.0000,1.1720)(10.0000,1.1008)(100.0000,1.1827)};

\addplot[color=red,  mark=x, line width=1pt] coordinates{(0.0100,0.5509)(0.1000,0.5508)(0.5000,0.6106)(1.0000,0.6800)(5.0000,1.0229)(10.0000,1.1138)(100.0000,1.1910)};

\addplot[color=green,  mark=x, line width=1pt] coordinates{(0.0100,0.9178)(0.1000,0.8137)(0.5000,0.5568)(1.0000,0.3259)(5.0000,0.0836)(10.0000,0.1129)(100.0000,0.0612)};


\addplot[color=orange,  mark=x, line width=1pt] coordinates{(0.0100,434.5237)(0.1000,363.5836)(0.5000,1199.9524)(1.0000,700.7765)(5.0000,200.6854)(10.0000,96.5869)(100.0000,118.8431)};

\addplot[name path=interleave_fully_connected_top,color=orange!70,draw=none] coordinates {(0.0100,1041.3797)(0.1000,865.8749)(0.5000,2889.7702)(1.0000,1682.4219)(5.0000,447.6599)(10.0000,222.3847)(100.0000,253.9018)};

\addplot[name path=interleave_fully_connected_down,color=orange!70,draw=none] coordinates {(0.0100,0.0001)(0.1000,0.0001)(0.5000,0.0001)(1.0000,0.0001)(5.0000,0.0001)(10.0000,0.0001)(100.0000,0.0001)};

\addplot[orange!50,fill opacity=0.1] fill between[of=interleave_fully_connected_top and interleave_fully_connected_down];

\end{axis}
\end{tikzpicture}
        \vspace{-2\baselineskip}
        \caption{$c_3=0.1$}
    \end{subfigure}
    \begin{subfigure}[b]{0.3\textwidth}
        \centering
        \begin{tikzpicture}

\begin{axis}[
    name=bottom axis,   
    width=6cm,
    height=6cm,
    xlabel=$c_2/c_1$,
    y label style={at={(0.0,1.0)}},
    xmin = 0.01, xmax = 100,     
    xtick={0.01, 0.1,1,10, 100}, 
    xticklabels={$0.01$, $0.1$, $1$, $10$, $100$}, 
    xmode=log,
    ymin = 0, ymax = 1.7,   
    ytick={0.0, 0.5, 1.0,1.5}, 
    yticklabels={-, $0.5$, $1.0$, $1.5$, $2.0$}, 
    grid = major,
    grid style=dashed,
    axis x line*=bottom,
    legend style={at={(2.7,1.3)},anchor=west}, 
]

\addplot[color=black,  mark=x, line width=1pt] coordinates{(0.0100,1.0000)(0.1000,1.0000)(0.5000,1.0000)(1.0000,1.0000)(5.0000,1.0000)(10.0000,1.0000)(100.0000,1.0000)};

\addplot[color=blue,  mark=x, line width=1pt] coordinates{(0.0100,0.6766)(0.1000,0.6646)(0.5000,0.6705)(1.0000,0.7248)(5.0000,0.9132)(10.0000,0.9469)(100.0000,1.1656)};

\addplot[color=red,  mark=x, line width=1pt] coordinates{(0.0100,0.4770)(0.1000,0.4730)(0.5000,0.5827)(1.0000,0.6705)(5.0000,0.9260)(10.0000,0.9911)(100.0000,1.1627)};

\addplot[color=green,  mark=x, line width=1pt] coordinates{(0.0100,0.9416)(0.1000,0.9056)(0.5000,0.7241)(1.0000,0.5161)(5.0000,0.1459)(10.0000,0.1450)(100.0000,0.0685)};


\addplot[name path=only_original_None_top,color=gray!70,draw=none] coordinates {(0.0100,1.0000)(0.1000,1.0000)(0.5000,1.0000)(1.0000,1.0000)(5.0000,1.0000)(10.0000,1.0000)(100.0000,1.0000)};

\addplot[name path=only_original_None_down,color=gray!70,draw=none] coordinates {(0.0100,1.0000)(0.1000,1.0000)(0.5000,1.0000)(1.0000,1.0000)(5.0000,1.0000)(10.0000,1.0000)(100.0000,1.0000)};

\addplot[gray!50,fill opacity=0.1] fill between[of=only_original_None_top and only_original_None_down];

\addplot[name path=interleave_cayley_top,color=blue!70,draw=none] coordinates {(0.0100,0.9560)(0.1000,0.9654)(0.5000,0.9888)(1.0000,0.9630)(5.0000,1.0708)(10.0000,1.1429)(100.0000,1.4334)};

\addplot[name path=interleave_cayley_down,color=blue!70,draw=none] coordinates {(0.0100,0.3972)(0.1000,0.3637)(0.5000,0.3522)(1.0000,0.4865)(5.0000,0.7557)(10.0000,0.7508)(100.0000,0.8979)};

\addplot[blue!50,fill opacity=0.1] fill between[of=interleave_cayley_top and interleave_cayley_down];

\addplot[name path=interleave_aligned_cayley_top,color=red!70,draw=none] coordinates {(0.0100,0.5897)(0.1000,0.6196)(0.5000,0.7065)(1.0000,0.7716)(5.0000,1.0812)(10.0000,1.2022)(100.0000,1.3911)};

\addplot[name path=interleave_aligned_cayley_down,color=red!70,draw=none] coordinates {(0.0100,0.3642)(0.1000,0.3264)(0.5000,0.4588)(1.0000,0.5694)(5.0000,0.7708)(10.0000,0.7800)(100.0000,0.9343)};

\addplot[red!50,fill opacity=0.1] fill between[of=interleave_aligned_cayley_top and interleave_aligned_cayley_down];

\addplot[name path=interleave_distance_d_pairs_top,color=green!70,draw=none] coordinates {(0.0100,1.0462)(0.1000,1.0715)(0.5000,0.9079)(1.0000,0.7060)(5.0000,0.1974)(10.0000,0.1951)(100.0000,0.0761)};

\addplot[name path=interleave_distance_d_pairs_down,color=green!70,draw=none] coordinates {(0.0100,0.8370)(0.1000,0.7396)(0.5000,0.5403)(1.0000,0.3262)(5.0000,0.0945)(10.0000,0.0948)(100.0000,0.0610)};

\addplot[green!50,fill opacity=0.1] fill between[of=interleave_distance_d_pairs_top and interleave_distance_d_pairs_down];




\end{axis}

\begin{axis}[
    at=(bottom axis.north),
    anchor=south, yshift=\pgfkeysvalueof{/tikz/axis break gap},
    width=6cm,
    height=3cm,
    xmin = 0.01, xmax = 100,     
    xtick={0.01, 0.1,1,10, 100}, 
    xticklabels={$0.01$, $0.1$, $1$, $10$, $100$}, 
    xmode=log,
    ymin = 20, ymax = 2000,   
    ytick={10, 100, 1000},          
    ymode=log,
    grid = major,
    grid style=dashed,
    title style={yshift=-1ex, text centered},  
    axis x line*=top,
    legend pos=south east,
    legend cell align=left,
    xticklabel=\empty,
    legend style={font=\footnotesize},
    legend pos=north east,
    after end axis/.code={
         \draw (rel axis cs:0,0) +(-2mm,-1mm) -- +(2mm,1mm)
              ++(0pt,-\pgfkeysvalueof{/tikz/axis break gap})
              +(-2mm,-1mm) -- +(2mm,1mm)
              (rel axis cs:0,0) +(0mm,0mm) -- +(0mm,0mm)
              ++(0pt,-\pgfkeysvalueof{/tikz/axis break gap})
              +(-2mm,-1mm) -- +(2mm,1mm);
}]


\addplot[color=black,  mark=x, line width=1pt] coordinates{(0.0100,1.0000)(0.1000,1.0000)(0.5000,1.0000)(1.0000,1.0000)(5.0000,1.0000)(10.0000,1.0000)(100.0000,1.0000)};

\addplot[color=blue,  mark=x, line width=1pt] coordinates{(0.0100,0.8419)(0.1000,0.8808)(0.5000,0.8185)(1.0000,0.8854)(5.0000,1.1720)(10.0000,1.1008)(100.0000,1.1827)};

\addplot[color=red,  mark=x, line width=1pt] coordinates{(0.0100,0.5509)(0.1000,0.5508)(0.5000,0.6106)(1.0000,0.6800)(5.0000,1.0229)(10.0000,1.1138)(100.0000,1.1910)};

\addplot[color=green,  mark=x, line width=1pt] coordinates{(0.0100,0.9178)(0.1000,0.8137)(0.5000,0.5568)(1.0000,0.3259)(5.0000,0.0836)(10.0000,0.1129)(100.0000,0.0612)};


\addplot[color=orange,  mark=x, line width=1pt] coordinates{(0.0100,343.3819)(0.1000,205.2094)(0.5000,217.9990)(1.0000,589.6864)(5.0000,285.7007)(10.0000,176.5057)(100.0000,72.8793)};

\addplot[name path=interleave_fully_connected_top,color=orange!70,draw=none] coordinates {(0.0100,826.5767)(0.1000,480.7775)(0.5000,523.9405)(1.0000,1419.6826)(5.0000,677.5498)(10.0000,420.8065)(100.0000,159.9141)};

\addplot[name path=interleave_fully_connected_down,color=orange!70,draw=none] coordinates {(0.0100,0.0001)(0.1000,0.0001)(0.5000,0.0001)(1.0000,0.0001)(5.0000,0.0001)(10.0000,0.0001)(100.0000,0.0001)};

\addplot[orange!50,fill opacity=0.1] fill between[of=interleave_fully_connected_top and interleave_fully_connected_down];

\end{axis}
\end{tikzpicture}
        \vspace{-2\baselineskip}
        \caption{$c_3=0.2$}
    \end{subfigure}
    \end{small}
    \caption{A sweep of $c_2$/$c_1$ for three values of $c_3$ in Data A, fixing $c_1=1.0$. \texttt{base-graph-only} is the baseline with no rewiring. Each other colour denotes a rewirer introduced by interleaving with the base graph across GNN layers.}
    \label{fig:salientdists-sweep}
\end{figure*}

Figure~\ref{fig:salientdists-sweep} shows sweeps over over $c_2/c_1$ for this dataset, at three values of $c_3$.

At $c_3 = 0$, standard message passing in \texttt{base-graph-only} outperforms both \texttt{cayley} and \texttt{aligned-cayley}. We surmise that the base graph is well-suited to a ``distance counting'' mechanism and hence capable of learning the distance $d$ interactions in theory. 
Interleaving with rewirers corrupts this mechanism and impedes performance, while getting no wins from improving average commute times since $c_3=0$. 

For $c_3 \neq 0$ and $c_2 < c_1$, we observe that \texttt{cayley} and \texttt{aligned-cayley} outperform the no-rewiring baseline.
We argue that this regime is the most interesting regime for realistic use cases. It has weak but nonzero interactions between all nodes ($c_3 \neq 0$), and moderate-strength special interactions that are weaker than 1-hop interactions on the base graph ($c_2 < c_1$). 
For example, at $c_3 = 0.2$ and $c_2 = 0.1$, the Cayley expander reduces MSEs to 70\% and the aligned Cayley to 50\% of the no-rewiring baseline. 
We believe \texttt{aligned-cayley} performs better than \texttt{cayley} on average because it captures a larger proportion of distance-$d$ edges. 
As $c_3$ increases,
the performance gap between \texttt{aligned-cayley} and \texttt{cayley} shrinks, because the relative importance of distance-$d$ compared to global interactions weakens.

\texttt{distance-d-pairs} is a tempting rewiring solution. Results show that this rewirer allows the model to learn almost perfectly when $c_2$ is large -- that is, when the special expert-provided priors dominate the type of interactions in the graph. 
However, this naive solution is suboptimal in other regimes. 
Namely, when $c_3 > 0$ and $c_1$ dominates $c_2$, \texttt{aligned-cayley} performs best by facilitating communication between all nodes while favouring distance-$d$ interactions through our prior-informed alignment scheme.

Finally, \texttt{fully-connected} achieves the worst MSEs by several orders of magnitude across all tested parameter settings. 
We suspect the large number of redundant messages is responsible for this method's poor OOD generalisation.

\paragraph{Data B}

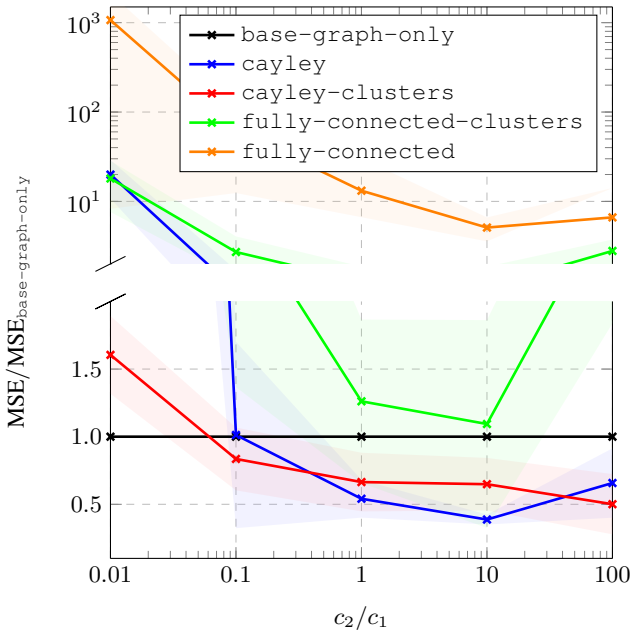
\begin{figure}[htb]
\centering
\vspace{1.0\baselineskip}
\begin{small}
\begin{tikzpicture}
\begin{axis}[
    name=bottom axis,   
    legend cell align=left,
    width=\linewidth,
    height=5cm,
    ylabel=$\textrm{MSE}/\textrm{MSE}_{\texttt{base-graph-only}}$,
    xlabel=$c_2/c_1$,
    y label style={at={(0.0,1.0)}},
    xmin = 0.01, xmax = 100,     
    xtick={0.01, 0.1,1,10, 100}, 
    xticklabels={$0.01$, $0.1$, $1$, $10$, $100$}, 
    xmode=log,
    ymin = 0.1, ymax = 2,   
    ytick={0.0, 0.5, 1.0,1.5}, 
    yticklabels={-, $0.5$, $1.0$, $1.5$, $2.0$}, 
    grid = major,
    grid style=dashed,
    axis x line*=bottom,
    legend pos=south east,
]
\addplot[color=black,  mark=x, line width=1pt] coordinates{(0.0100,1.0000)(0.1000,1.0000)(1.0000,1.0000)(10.0000,1.0000)(100.0000,1.0000)};

\addplot[color=blue,  mark=x, line width=1pt] coordinates{(0.0100,19.9572)(0.1000,1.0127)(1.0000,0.5411)(10.0000,0.3873)(100.0000,0.6574)};

\addplot[color=red,  mark=x, line width=1pt] coordinates{(0.0100,1.6050)(0.1000,0.8354)(1.0000,0.6644)(10.0000,0.6486)(100.0000,0.5007)};

\addplot[color=green,  mark=x, line width=1pt] coordinates{(0.0100,18.0465)(0.1000,2.7137)(1.0000,1.2616)(10.0000,1.0937)(100.0000,2.7915)};


\addplot[name path=only_original_None_top,color=gray!70,draw=none] coordinates {(0.0100,1.0000)(0.1000,1.0000)(1.0000,1.0000)(10.0000,1.0000)(100.0000,1.0000)};

\addplot[name path=only_original_None_down,color=gray!70,draw=none] coordinates {(0.0100,1.0000)(0.1000,1.0000)(1.0000,1.0000)(10.0000,1.0000)(100.0000,1.0000)};

\addplot[gray!50,fill opacity=0.1] fill between[of=only_original_None_top and only_original_None_down];

\addplot[name path=interleave_cayley_top,color=blue!70,draw=none] coordinates {(0.0100,29.1614)(0.1000,1.7019)(1.0000,0.6810)(10.0000,0.4241)(100.0000,0.9141)};

\addplot[name path=interleave_cayley_down,color=blue!70,draw=none] coordinates {(0.0100,10.7530)(0.1000,0.3234)(1.0000,0.4013)(10.0000,0.3505)(100.0000,0.4006)};

\addplot[blue!50,fill opacity=0.1] fill between[of=interleave_cayley_top and interleave_cayley_down];

\addplot[name path=interleave_unconnected_cayley_clusters_top,color=red!70,draw=none] coordinates {(0.0100,1.8928)(0.1000,1.0679)(1.0000,0.8814)(10.0000,0.8432)(100.0000,0.7237)};

\addplot[name path=interleave_unconnected_cayley_clusters_down,color=red!70,draw=none] coordinates {(0.0100,1.3172)(0.1000,0.6029)(1.0000,0.4475)(10.0000,0.4540)(100.0000,0.2776)};

\addplot[red!50,fill opacity=0.1] fill between[of=interleave_unconnected_cayley_clusters_top and interleave_unconnected_cayley_clusters_down];

\addplot[name path=interleave_fully_connected_clusters_top,color=green!70,draw=none] coordinates {(0.0100,28.4819)(0.1000,4.0607)(1.0000,1.8664)(10.0000,1.8624)(100.0000,3.7454)};

\addplot[name path=interleave_fully_connected_clusters_down,color=green!70,draw=none] coordinates {(0.0100,7.6110)(0.1000,1.3666)(1.0000,0.6569)(10.0000,0.3250)(100.0000,1.8376)};

\addplot[green!50,fill opacity=0.1] fill between[of=interleave_fully_connected_clusters_top and interleave_fully_connected_clusters_down];




\end{axis}

\begin{axis}[
    at=(bottom axis.north),
    anchor=south, yshift=\pgfkeysvalueof{/tikz/axis break gap},
    width=\linewidth,
    height=5cm,
    xmin = 0.01, xmax = 100,     
    xtick={0.01, 0.1,1,10, 100}, 
    xticklabels={$0.01$, $0.1$, $1$, $10$, $100$}, 
    xmode=log,
    ymin = 2, ymax = 1500,   
    ytick={10, 100, 1000},          
    ymode=log,
    grid = major,
    grid style=dashed,
    title style={yshift=-1ex, text centered},  
    axis x line*=top,
    legend pos=south east,
    legend cell align=left,
    xticklabel=\empty,
    legend style={font=\footnotesize},
    legend pos=north east,
    after end axis/.code={
         \draw (rel axis cs:0,0) +(-2mm,-1mm) -- +(2mm,1mm)
              ++(0pt,-\pgfkeysvalueof{/tikz/axis break gap})
              +(-2mm,-1mm) -- +(2mm,1mm)
              (rel axis cs:0,0) +(0mm,0mm) -- +(0mm,0mm)
              ++(0pt,-\pgfkeysvalueof{/tikz/axis break gap})
              +(-2mm,-1mm) -- +(2mm,1mm);
    }]

\addplot[color=black,  mark=x, line width=1pt] coordinates{(0.0100,1.0000)(0.1000,1.0000)(1.0000,1.0000)(10.0000,1.0000)(100.0000,1.0000)};

\addplot[color=blue,  mark=x, line width=1pt] coordinates{(0.0100,19.9572)(0.1000,1.0127)(1.0000,0.5411)(10.0000,0.3873)(100.0000,0.6574)};

\addplot[color=red,  mark=x, line width=1pt] coordinates{(0.0100,1.6050)(0.1000,0.8354)(1.0000,0.6644)(10.0000,0.6486)(100.0000,0.5007)};

\addplot[color=green,  mark=x, line width=1pt] coordinates{(0.0100,18.0465)(0.1000,2.7137)(1.0000,1.2616)(10.0000,1.0937)(100.0000,2.7915)};

\addplot[color=orange,  mark=x, line width=1pt] coordinates{(0.0100,1068.6309)(0.1000,62.6480)(1.0000,13.1745)(10.0000,5.0960)(100.0000,6.6263)};

\addplot[name path=only_original_None_top,color=gray!70,draw=none] coordinates {(0.0100,1.0000)(0.1000,1.0000)(1.0000,1.0000)(10.0000,1.0000)(100.0000,1.0000)};

\addplot[name path=only_original_None_down,color=gray!70,draw=none] coordinates {(0.0100,1.0000)(0.1000,1.0000)(1.0000,1.0000)(10.0000,1.0000)(100.0000,1.0000)};

\addplot[gray!50,fill opacity=0.1] fill between[of=only_original_None_top and only_original_None_down];

\addplot[name path=interleave_cayley_top,color=blue!70,draw=none] coordinates {(0.0100,29.1614)(0.1000,1.7019)(1.0000,0.6810)(10.0000,0.4241)(100.0000,0.9141)};

\addplot[name path=interleave_cayley_down,color=blue!70,draw=none] coordinates {(0.0100,10.7530)(0.1000,0.3234)(1.0000,0.4013)(10.0000,0.3505)(100.0000,0.4006)};

\addplot[blue!50,fill opacity=0.1] fill between[of=interleave_cayley_top and interleave_cayley_down];

\addplot[name path=interleave_unconnected_cayley_clusters_top,color=red!70,draw=none] coordinates {(0.0100,1.8928)(0.1000,1.0679)(1.0000,0.8814)(10.0000,0.8432)(100.0000,0.7237)};

\addplot[name path=interleave_unconnected_cayley_clusters_down,color=red!70,draw=none] coordinates {(0.0100,1.3172)(0.1000,0.6029)(1.0000,0.4475)(10.0000,0.4540)(100.0000,0.2776)};

\addplot[red!50,fill opacity=0.1] fill between[of=interleave_unconnected_cayley_clusters_top and interleave_unconnected_cayley_clusters_down];

\addplot[name path=interleave_fully_connected_clusters_top,color=green!70,draw=none] coordinates {(0.0100,28.4819)(0.1000,4.0607)(1.0000,1.8664)(10.0000,1.8624)(100.0000,3.7454)};

\addplot[name path=interleave_fully_connected_clusters_down,color=green!70,draw=none] coordinates {(0.0100,7.6110)(0.1000,1.3666)(1.0000,0.6569)(10.0000,0.3250)(100.0000,1.8376)};

\addplot[green!50,fill opacity=0.1] fill between[of=interleave_fully_connected_clusters_top and interleave_fully_connected_clusters_down];

\addplot[name path=interleave_fully_connected_top,color=orange!70,draw=none] coordinates {(0.0100,2128.9881)(0.1000,112.8937)(1.0000,30.2581)(10.0000,6.6216)(100.0000,14.1270)};

\addplot[name path=interleave_fully_connected_down,color=orange!70,draw=none] coordinates {(0.0100,8.2736)(0.1000,12.4023)(1.0000,0.0)(10.0000,3.5705)(100.0000,0.0)};

\addplot[orange!50,fill opacity=0.1] fill between[of=interleave_fully_connected_top and interleave_fully_connected_down];

\legend{
\texttt{base-graph-only}, 
\texttt{cayley},
\texttt{cayley-clusters},
\texttt{fully-connected-clusters}, 
\texttt{fully-connected},
}

\end{axis}

\end{tikzpicture}
\end{small}
\vspace{-2\baselineskip}
\caption{A sweep of $c_2/c_1$ in Data B, fixing $c_1+c_2=1$ (since the target grows very quickly for large $c_2$). \texttt{base-graph-only} is the baseline with no rewiring. Each other colour denotes a rewirer introduced by interleaving with the base graph across GNN layers.}
\label{fig:colourinteract-sweep}
\end{figure}

Figure~\ref{fig:colourinteract-sweep} 
shows a sweep over $c_2/c_1$ for this dataset.

Consider the low $c_2/c_1$ regime, where nearest-neighbour interactions dominate.
At $c_2/c_1=0.01$, the interleaved graphs and one-hot colour encodings are distractions.
As expected, \texttt{base-graph-only} is here the best approach.
Interleaving with \texttt{cayley}, \texttt{fully-connected-clusters}, or \texttt{fully-connected} results in worse MSEs by more than an order of magnitude.
In contrast, \texttt{cayley-clusters} is relatively robust to the distracting colour encodings, though still suboptimal.
We summarise this regime as follows:
the prior correctly identifies existence of node colours, but they are irrelevant to the task.
\texttt{cayley-clusters} learns this much better than the other rewiring solutions.

For moderate $c_2/c_1$, where both nearest-neighbour and colour interactions are important,
we find that interleaving with either Cayley-style expander significantly outperforms \texttt{base-graph-only}.
\texttt{cayley} and \texttt{cayley-clusters} are near-indistinguishable in this regime: 
the former achieves marginally lower MSEs and the latter marginally smaller variance.
This result is intuitive, as \texttt{base-graph-only} cannot capture all colour interactions, while the expanders make these colour connections accessible.
Morever, we again find that sparse Cayley expanders are a better solution than dense cliques:
\texttt{fully-connected} continues to perform very poorly,
and \texttt{fully-connected-clusters} is hugely unstable while underperforming the baseline on average.

When $c_2/c_1$ is large, colour interactions dominate.
At $c_2/c_1=100$, \texttt{cayley-clusters} outperforms all the other approaches on average, with \texttt{cayley} following closely behind.
The Cayley rewirers outperform \texttt{base-graph-only}, which can only do well when same-colour nodes are close together on the base graph by chance.
On the other hand, both fully-connected rewirers underperform \texttt{base-graph-only} by an order of magnitude.
This is a particularly surprising result in the case of \texttt{fully-connected-clusters}, because this rewiring contains all node pairs that interact.
The underperformance is reminiscent of the more general OOD issues in transformers, as discussed in Section~\ref{sec:egp}.
The largest cluster in \texttt{fully-conected-clusters} connects uncoloured nodes, representing our prior that such nodes should not interact.
In our setup, the size of this cluster grows linearly (i.e. graph size minus number of coloured nodes) with the total number of nodes.
This offers an explanation for the otherwise surprisingly poor performance of \texttt{fully-conected-clusters}, and further motivates our sparse rewiring approaches.

In summary, for this dataset, our \texttt{cayley-clusters} is consistently competitive with or better than other interleaving approaches.
Meanwhile, it has consistently lower variance, lending the approach well to practical application where one has less certainty on the prior.

\section{Case Study}

The \texttt{cayley-clusters} rewiring aims to interact similar groups of nodes as much as possible. 
We now apply this rewiring approach to a real-world dataset, \texttt{ogbn-arxiv}.
The task is to tranductively classify the subject area of papers in a citation network \cite{huopen2020}. 
We chose this dataset since citation graphs tend to exhibit homophily, making them well-suited for \texttt{cayley-clusters}.
Additionally, \texttt{ogbn-arxiv} is sufficiently challenging that simpler flavours of GNNs models have not already oversaturated performance, in contrast to older citation graphs like Cora and PubMed.

We begin with a sanity check: can we boost performance by directly using ground truth class labels as a binary similarity measure?
We compare \texttt{base-graph-only} and \texttt{cayley} with an interleaved \texttt{cayley-clusters} rewiring, where colours are assigned by the labels for all (train, val, and test) nodes.
Though this constitutes data leakage, we use it to illustrate the potential performance improvement offered by our rewiring in the extreme case of a perfect prior.
Additionally, we note that a fully-connected rewiring is computationally intractable; our rewiring is sparse.

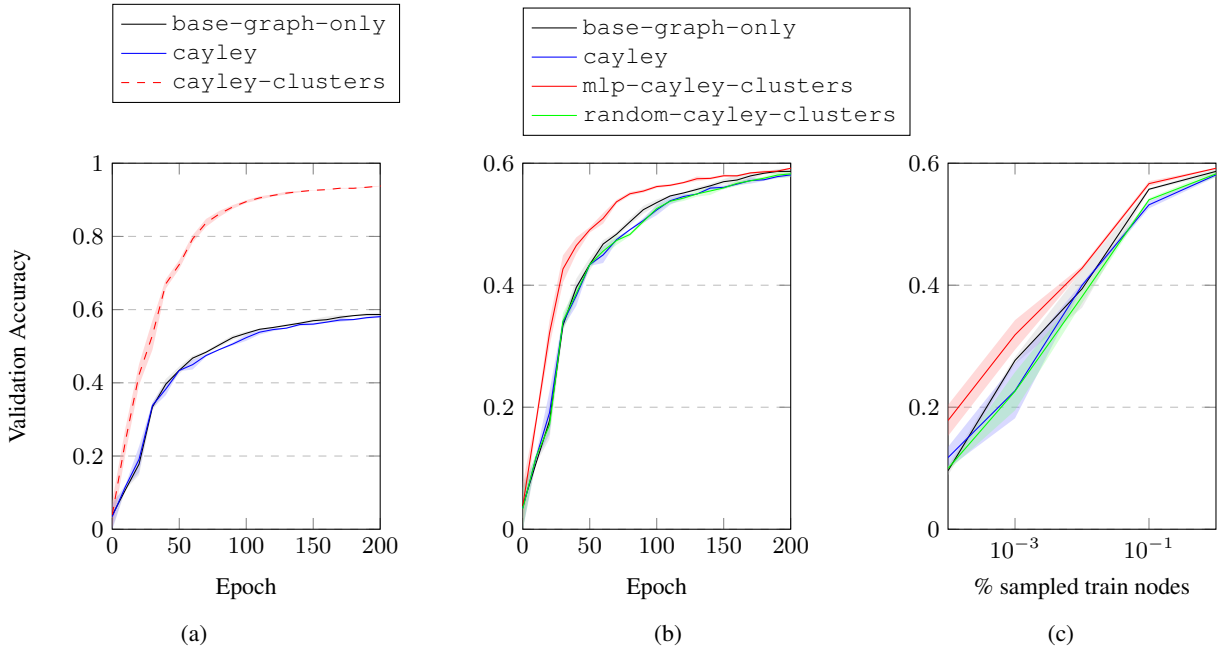
\begin{figure*}
\begin{small}
    \centering
    \begin{subfigure}[t]{0.3\textwidth}
        \centering
        \begin{tikzpicture}
\begin{axis}[
    xmin=0.0, xmax=200,
    ymin=0, ymax=1,
    ymajorgrids=true,
    xlabel=Epoch,
    ylabel=Validation Accuracy,
    grid style=dashed,
    width=\textwidth,
    height=2.54in,
    legend style={at={(0,1.3)},anchor=west},
    legend cell align=left,
]
\addplot[color=black, mark=none] coordinates{(0, 0.0383)  (10, 0.1089)  (20, 0.178)  (30, 0.3342)  (40, 0.3965)  (50, 0.4341)  (60, 0.4675)  (70, 0.4836)  (80, 0.5039)  (90, 0.5238)  (100, 0.5354)  (110, 0.5463)  (120, 0.5512)  (130, 0.557)  (140, 0.5629)  (150, 0.5697)  (160, 0.5725)  (170, 0.5792)  (180, 0.5835)  (190, 0.5865)  (200, 0.5866)  (210, 0.5866)};

\addplot[color=blue, mark=none] coordinates{(0, 0.0346)  (10, 0.1178)  (20, 0.1911)  (30, 0.3383)  (40, 0.383)  (50, 0.4337)  (60, 0.4497)  (70, 0.4746)  (80, 0.491)  (90, 0.5061)  (100, 0.5234)  (110, 0.5384)  (120, 0.5454)  (130, 0.5494)  (140, 0.5592)  (150, 0.5604)  (160, 0.5658)  (170, 0.5716)  (180, 0.5728)  (190, 0.578)  (200, 0.5808)  (210, 0.5808)};

\addplot[color=red, mark=none, dashed] coordinates{(0, 0.0398)  (10, 0.2442)  (20, 0.4223)  (30, 0.5302)  (40, 0.6706)  (50, 0.7231)  (60, 0.7933)  (70, 0.8369)  (80, 0.8616)  (90, 0.8805)  (100, 0.895)  (110, 0.9052)  (120, 0.912)  (130, 0.9183)  (140, 0.9225)  (150, 0.926)  (160, 0.9277)  (170, 0.9316)  (180, 0.9315)  (190, 0.9343)  (200, 0.9376)  (210, 0.9376)};

\addplot[name path=class_all_top,color=gray!70,draw=none] coordinates {(0, 0.0785)  (10, 0.1096)  (20, 0.2079)  (30, 0.3444)  (40, 0.4108)  (50, 0.4446)  (60, 0.4739)  (70, 0.4874)  (80, 0.5096)  (90, 0.5301)  (100, 0.5405)  (110, 0.5478)  (120, 0.5524)  (130, 0.5604)  (140, 0.5662)  (150, 0.574)  (160, 0.5812)  (170, 0.5852)  (180, 0.585)  (190, 0.5906)  (200, 0.5913)  (210, 0.5913)};

\addplot[name path=class_all_down,color=gray!70,draw=none] coordinates {(0, -0.0018)  (10, 0.1082)  (20, 0.148)  (30, 0.324)  (40, 0.3821)  (50, 0.4236)  (60, 0.4611)  (70, 0.4798)  (80, 0.4981)  (90, 0.5174)  (100, 0.5304)  (110, 0.5449)  (120, 0.5501)  (130, 0.5537)  (140, 0.5596)  (150, 0.5655)  (160, 0.5638)  (170, 0.5731)  (180, 0.5819)  (190, 0.5824)  (200, 0.5819)  (210, 0.5819)};

\addplot[gray!50,fill opacity=0.3] fill between[of=class_all_top and class_all_down];

\addplot[name path=class_all_top,color=blue!70,draw=none] coordinates {(0, 0.0689)  (10, 0.1228)  (20, 0.2266)  (30, 0.3462)  (40, 0.3998)  (50, 0.4358)  (60, 0.4631)  (70, 0.4782)  (80, 0.4922)  (90, 0.5101)  (100, 0.5319)  (110, 0.5449)  (120, 0.5496)  (130, 0.5501)  (140, 0.5628)  (150, 0.5634)  (160, 0.5676)  (170, 0.5768)  (180, 0.5745)  (190, 0.5807)  (200, 0.5842)  (210, 0.5842)};

\addplot[name path=class_all_down,color=blue!70,draw=none] coordinates {(0, 0.0003)  (10, 0.1128)  (20, 0.1556)  (30, 0.3304)  (40, 0.3663)  (50, 0.4316)  (60, 0.4364)  (70, 0.4711)  (80, 0.4898)  (90, 0.5022)  (100, 0.5148)  (110, 0.5319)  (120, 0.5412)  (130, 0.5487)  (140, 0.5556)  (150, 0.5574)  (160, 0.564)  (170, 0.5664)  (180, 0.571)  (190, 0.5753)  (200, 0.5775)  (210, 0.5775)};

\addplot[blue!50,fill opacity=0.3] fill between[of=class_all_top and class_all_down];

\addplot[name path=class_all_top,color=red!70,draw=none] coordinates {(0, 0.077)  (10, 0.2795)  (20, 0.4512)  (30, 0.5705)  (40, 0.6836)  (50, 0.7351)  (60, 0.8034)  (70, 0.8477)  (80, 0.8698)  (90, 0.8856)  (100, 0.899)  (110, 0.9096)  (120, 0.915)  (130, 0.922)  (140, 0.9245)  (150, 0.9264)  (160, 0.9282)  (170, 0.9327)  (180, 0.9323)  (190, 0.9358)  (200, 0.9383)  (210, 0.9383)};

\addplot[name path=class_all_down,color=red!70,draw=none] coordinates {(0, 0.0026)  (10, 0.209)  (20, 0.3934)  (30, 0.49)  (40, 0.6577)  (50, 0.711)  (60, 0.7832)  (70, 0.8262)  (80, 0.8534)  (90, 0.8755)  (100, 0.891)  (110, 0.9007)  (120, 0.909)  (130, 0.9146)  (140, 0.9205)  (150, 0.9255)  (160, 0.9273)  (170, 0.9304)  (180, 0.9307)  (190, 0.9328)  (200, 0.9369)  (210, 0.9369)};

\addplot[red!50,fill opacity=0.3] fill between[of=class_all_top and class_all_down];

\legend{\texttt{base-graph-only}, \texttt{cayley}, \texttt{cayley-clusters}}
\end{axis}
\end{tikzpicture}
        \vspace{-\baselineskip}
        \caption{}
        \label{fig:arxiv-sanity-check}
    \end{subfigure}
    \hspace{1cm}
    \begin{subfigure}[t]{0.3\textwidth}
        \centering
        \begin{tikzpicture}
\centering
\begin{axis}[
    xmin=0.0, xmax=200,
    ymin=0, ymax=0.6,
    ymajorgrids=true,
    xlabel=Epoch,
    grid style=dashed,
    width=\textwidth,
    height=2.54in,
    legend style={at={(0,1.25)},anchor=west},
    legend cell align=left,
]
\addplot[color=black, mark=none] coordinates{(0, 0.0383)  (10, 0.1089)  (20, 0.178)  (30, 0.3342)  (40, 0.3965)  (50, 0.4341)  (60, 0.4675)  (70, 0.4836)  (80, 0.5039)  (90, 0.5238)  (100, 0.5354)  (110, 0.5463)  (120, 0.5512)  (130, 0.557)  (140, 0.5629)  (150, 0.5697)  (160, 0.5725)  (170, 0.5792)  (180, 0.5835)  (190, 0.5865)  (200, 0.5866)  (210, 0.5866)};

\addplot[color=blue, mark=none] coordinates{(0, 0.0346)  (10, 0.1178)  (20, 0.1911)  (30, 0.3383)  (40, 0.383)  (50, 0.4337)  (60, 0.4497)  (70, 0.4746)  (80, 0.491)  (90, 0.5061)  (100, 0.5234)  (110, 0.5384)  (120, 0.5454)  (130, 0.5494)  (140, 0.5592)  (150, 0.5604)  (160, 0.5658)  (170, 0.5716)  (180, 0.5728)  (190, 0.578)  (200, 0.5808)  (210, 0.5808)};

\addplot[color=red, mark=none] coordinates{(0, 0.0384)  (10, 0.1773)  (20, 0.3216)  (30, 0.4266)  (40, 0.4651)  (50, 0.4908)  (60, 0.5096)  (70, 0.5367)  (80, 0.5499)  (90, 0.5543)  (100, 0.5617)  (110, 0.5638)  (120, 0.5682)  (130, 0.5745)  (140, 0.5751)  (150, 0.5794)  (160, 0.5793)  (170, 0.5841)  (180, 0.5858)  (190, 0.587)  (200, 0.5917)  (210, 0.5917)};

\addplot[color=green, mark=none] coordinates{(0, 0.0337)  (10, 0.1192)  (20, 0.1709)  (30, 0.3411)  (40, 0.3867)  (50, 0.4339)  (60, 0.4558)  (70, 0.4733)  (80, 0.4832)  (90, 0.5046)  (100, 0.5256)  (110, 0.5376)  (120, 0.5428)  (130, 0.5496)  (140, 0.5545)  (150, 0.5595)  (160, 0.567)  (170, 0.5712)  (180, 0.5754)  (190, 0.5805)  (200, 0.5825)  (210, 0.5825)};

\addplot[name path=random_top,color=gray!70,draw=none] coordinates {(0, 0.0785)  (10, 0.1096)  (20, 0.2079)  (30, 0.3444)  (40, 0.4108)  (50, 0.4446)  (60, 0.4739)  (70, 0.4874)  (80, 0.5096)  (90, 0.5301)  (100, 0.5405)  (110, 0.5478)  (120, 0.5524)  (130, 0.5604)  (140, 0.5662)  (150, 0.574)  (160, 0.5812)  (170, 0.5852)  (180, 0.585)  (190, 0.5906)  (200, 0.5913)  (210, 0.5913)};

\addplot[name path=random_down,color=gray!70,draw=none] coordinates {(0, -0.0018)  (10, 0.1082)  (20, 0.148)  (30, 0.324)  (40, 0.3821)  (50, 0.4236)  (60, 0.4611)  (70, 0.4798)  (80, 0.4981)  (90, 0.5174)  (100, 0.5304)  (110, 0.5449)  (120, 0.5501)  (130, 0.5537)  (140, 0.5596)  (150, 0.5655)  (160, 0.5638)  (170, 0.5731)  (180, 0.5819)  (190, 0.5824)  (200, 0.5819)  (210, 0.5819)};

\addplot[gray!50,fill opacity=0.3] fill between[of=random_top and random_down];

\addplot[name path=random_top,color=blue!70,draw=none] coordinates {(0, 0.0689)  (10, 0.1228)  (20, 0.2266)  (30, 0.3462)  (40, 0.3998)  (50, 0.4358)  (60, 0.4631)  (70, 0.4782)  (80, 0.4922)  (90, 0.5101)  (100, 0.5319)  (110, 0.5449)  (120, 0.5496)  (130, 0.5501)  (140, 0.5628)  (150, 0.5634)  (160, 0.5676)  (170, 0.5768)  (180, 0.5745)  (190, 0.5807)  (200, 0.5842)  (210, 0.5842)};

\addplot[name path=random_down,color=blue!70,draw=none] coordinates {(0, 0.0003)  (10, 0.1128)  (20, 0.1556)  (30, 0.3304)  (40, 0.3663)  (50, 0.4316)  (60, 0.4364)  (70, 0.4711)  (80, 0.4898)  (90, 0.5022)  (100, 0.5148)  (110, 0.5319)  (120, 0.5412)  (130, 0.5487)  (140, 0.5556)  (150, 0.5574)  (160, 0.564)  (170, 0.5664)  (180, 0.571)  (190, 0.5753)  (200, 0.5775)  (210, 0.5775)};

\addplot[blue!50,fill opacity=0.3] fill between[of=random_top and random_down];

\addplot[name path=random_top,color=red!70,draw=none] coordinates {(0, 0.0749)  (10, 0.1859)  (20, 0.3425)  (30, 0.4489)  (40, 0.4785)  (50, 0.4952)  (60, 0.5182)  (70, 0.54)  (80, 0.553)  (90, 0.5569)  (100, 0.564)  (110, 0.5657)  (120, 0.5699)  (130, 0.5782)  (140, 0.5772)  (150, 0.5814)  (160, 0.5806)  (170, 0.5852)  (180, 0.5881)  (190, 0.5886)  (200, 0.5927)  (210, 0.5927)};

\addplot[name path=random_down,color=red!70,draw=none] coordinates {(0, 0.002)  (10, 0.1687)  (20, 0.3007)  (30, 0.4043)  (40, 0.4517)  (50, 0.4863)  (60, 0.5009)  (70, 0.5334)  (80, 0.5469)  (90, 0.5517)  (100, 0.5593)  (110, 0.5619)  (120, 0.5666)  (130, 0.5709)  (140, 0.573)  (150, 0.5774)  (160, 0.578)  (170, 0.583)  (180, 0.5835)  (190, 0.5854)  (200, 0.5908)  (210, 0.5908)};

\addplot[red!50,fill opacity=0.3] fill between[of=random_top and random_down];

\addplot[name path=random_top,color=green!70,draw=none] coordinates {(0, 0.0676)  (10, 0.1237)  (20, 0.1825)  (30, 0.3546)  (40, 0.3964)  (50, 0.4384)  (60, 0.463)  (70, 0.4807)  (80, 0.4862)  (90, 0.5082)  (100, 0.5312)  (110, 0.5415)  (120, 0.5477)  (130, 0.554)  (140, 0.561)  (150, 0.5614)  (160, 0.5713)  (170, 0.5737)  (180, 0.5782)  (190, 0.5852)  (200, 0.5877)  (210, 0.5877)};

\addplot[name path=random_down,color=green!70,draw=none] coordinates {(0, -0.0001)  (10, 0.1148)  (20, 0.1593)  (30, 0.3276)  (40, 0.377)  (50, 0.4294)  (60, 0.4486)  (70, 0.466)  (80, 0.4803)  (90, 0.501)  (100, 0.5199)  (110, 0.5338)  (120, 0.538)  (130, 0.5452)  (140, 0.548)  (150, 0.5577)  (160, 0.5626)  (170, 0.5687)  (180, 0.5725)  (190, 0.5758)  (200, 0.5772)  (210, 0.5772)};

\addplot[green!50,fill opacity=0.3] fill between[of=random_top and random_down];

\legend{\texttt{base-graph-only}, \texttt{cayley}, \texttt{mlp-cayley-clusters}, \texttt{random-cayley-clusters}}
\end{axis}
\end{tikzpicture}
        \vspace{-\baselineskip}
        \caption{}
        \label{fig:ogbn_faster_convergence}
    \end{subfigure}
    \begin{subfigure}[t]{0.3\textwidth}
        \centering
        \begin{tikzpicture}
\begin{axis}[
    xmin=0.0001, xmax=1,
    ymin=0, ymax=0.6,
    ymajorgrids=true,
    xlabel=\% sampled train nodes,
    xmode=log,
    grid style=dashed,
    width=\textwidth,
    height=2.54in,
]
\addplot[color=black, mark=none] coordinates{(0.0001, 0.096)  (0.001, 0.2767)  (0.01, 0.3939)  (0.1, 0.5573)  (1.0, 0.5866)};

\addplot[color=blue, mark=none] coordinates{(0.0001, 0.1175)  (0.001, 0.2271)  (0.01, 0.3998)  (0.1, 0.5318)  (1.0, 0.5808)};

\addplot[color=red, mark=none] coordinates{(0.0001, 0.1783)  (0.001, 0.3186)  (0.01, 0.4278)  (0.1, 0.5665)  (1.0, 0.5917)};

\addplot[color=green, mark=none] coordinates{(0.0001, 0.0998)  (0.001, 0.2261)  (0.01, 0.3811)  (0.1, 0.54)  (1.0, 0.5825)};

\addplot[name path=None-only_original_top,color=gray!70,draw=none] coordinates {(0.0001, 0.1037)  (0.001, 0.2825)  (0.01, 0.4248)  (0.1, 0.559)  (1.0, 0.5913)};

\addplot[name path=None-only_original_down,color=gray!70,draw=none] coordinates {(0.0001, 0.0883)  (0.001, 0.2708)  (0.01, 0.363)  (0.1, 0.5555)  (1.0, 0.5819)};

\addplot[gray!50,fill opacity=0.3] fill between[of=None-only_original_top and None-only_original_down];

\addplot[name path=cayley-interleave_top,color=blue!70,draw=none] coordinates {(0.0001, 0.1351)  (0.001, 0.2724)  (0.01, 0.4034)  (0.1, 0.5376)  (1.0, 0.5842)};

\addplot[name path=cayley-interleave_down,color=blue!70,draw=none] coordinates {(0.0001, 0.0998)  (0.001, 0.1818)  (0.01, 0.3963)  (0.1, 0.526)  (1.0, 0.5775)};

\addplot[blue!50,fill opacity=0.3] fill between[of=cayley-interleave_top and cayley-interleave_down];

\addplot[name path=mlp_all-interleave_top,color=red!70,draw=none] coordinates {(0.0001, 0.2038)  (0.001, 0.3428)  (0.01, 0.4317)  (0.1, 0.5705)  (1.0, 0.5927)};

\addplot[name path=mlp_all-interleave_down,color=red!70,draw=none] coordinates {(0.0001, 0.1528)  (0.001, 0.2943)  (0.01, 0.4239)  (0.1, 0.5625)  (1.0, 0.5908)};

\addplot[red!50,fill opacity=0.3] fill between[of=mlp_all-interleave_top and mlp_all-interleave_down];

\addplot[name path=random-interleave_top,color=green!70,draw=none] coordinates {(0.0001, 0.1033)  (0.001, 0.2575)  (0.01, 0.3918)  (0.1, 0.5428)  (1.0, 0.5877)};

\addplot[name path=random-interleave_down,color=green!70,draw=none] coordinates {(0.0001, 0.0963)  (0.001, 0.1948)  (0.01, 0.3704)  (0.1, 0.5371)  (1.0, 0.5772)};

\addplot[green!50,fill opacity=0.3] fill between[of=random-interleave_top and random-interleave_down];

\end{axis}
\end{tikzpicture}
        \caption{}
        \label{fig:ogbn_subsample}
    \end{subfigure}
\end{small}
\caption{Results on \texttt{ogbn-arxiv}. 
(a) Validation accuracy vs. train epoch for the sanity check on \texttt{ogbn-arxiv}. Dashed lines represent cases where the ground truth labels are directly used to impute colours in the construction of the non-parametric rewirer. 
(b) Validation accuracy vs. train epoch shows that our \texttt{mlp-cayley-clusters} achieves faster convergence than baselines.
(c) Final validation accuracy as a function of subsampling fraction shows that a valuable prior provided through the \texttt{mlp-cayley-clusters} rewiring improves final accuracies.
Each line denotes the mean of three runs, with error bars
representing one standard deviation.
}
\label{fig:arxiv-plots}
\end{figure*}

Figure~\ref{fig:arxiv-sanity-check} shows the results of the sanity check.
Compared to the baselines, where train accuracy converges to 67\% and validation accuracy to 62\%, \texttt{cayley-clusters} achieves high-90s\% accuracy on train and validation nodes.
This is consistent with our understanding that the Cayley clusters efficiently reinforce homophilic connections across the entire graph.
Modifying the graph topology using ground truth labels,
but \textit{without} model parameter tuning on the validation labels, yields a massive performance boost.

We now turn to a more realistic scenario: without this perfect prior, can we extract a useful measure of similarity from node features? 
Inspired by \citet{deac2023evolving}, we train a weak MLP classifier on the train node features, ignoring the graph's connectivity.
Colours are then assigned based on the predictions of this MLP, leading to the \texttt{mlp-cayley-clusters} rewiring.
As controls, we also report the \texttt{cayley} baseline, and \texttt{random-cayley-clusters} where the colours are assigned randomly.
Figure~\ref{fig:ogbn_faster_convergence} shows that all methods reach comparable validation performance after approximately 200 epochs.
However, our \texttt{mlp-cayley-clusters} enjoys substantially faster convergence.
Notably, neither \texttt{cayley} nor \texttt{random-cayley-clusters} learns more quickly than the no-rewiring baseline.
This suggests that the effective incorporation of our homophily prior, rather than a prior-agnostic reduction in global commute times, is responsible for the improved convergence.

Finally, we want to probe the intermediate situation where the prior offers information not available as labels.
We simulate this effect by subsampling the train nodes when training the GNN, achieving a low-resourced setting where labels are sparse.
Since our weak MLP classifier learns from the entire train set, its corresponding \texttt{mlp-cayley-clusters} rewiring offers information not otherwise available to the GNN.
The more extreme the subsampling, the more ``information-rich'' the rewiring is.
Results are reported in Figure~\ref{fig:ogbn_subsample}.
As before, both \texttt{cayley} and \texttt{random-cayley-clusters} attain the same performance as the no-rewiring baseline.
In contrast, \texttt{mlp-cayley-clusters} achieves statistically significant improvement to final validation accuracies when label data is sparse.
This experiment shows that incorporating a prior is valuable in the low-data regime.
As the label availability improves, the advantage of \texttt{mlp-cayley-clusters} narrows.
Meanwhile, the \texttt{random-cayley-clusters} baseline demonstrates that if the prior turns out to be incorrect, our approach does not hurt performance.

For all experiments in this section, we used a GIN model with 3 layers, 128 hidden channels, 0.5 dropout probability, and batch norm. 
The model is trained for 200 epochs with 0.001 learning rate.
We do not perform any hyperparameter tuning.

\section{Conclusion}

We began with the question: Given an expert prior on which node pairs need to interact in a graph learning problem, how can we optimise commute time between said pairs?
We constructed two synthetic datasets that schematise two realistic graph priors and proposed non-parametric rewiring methods tailored to these priors. 
We investigated the regimes in which our graph rewiring methods outperform base-graph-only and Cayley expander baselines. 
In our case study, we found using a prior with our bespoke rewiring method leads to faster convergence on \texttt{ogbn-arxiv} and performance gains in low-data regimes.

\section*{Acknowledgements}

We thank our anonymous reviewers for their comments, and Alex Vitvitskyi and Simon Osindero for reviewing the paper prior to submission.

\bibliography{main}
\bibliographystyle{icml2024}

\end{document}